\documentclass[unpublished]{JHEP3}

\usepackage{epsfig,multicol,bbm,amsmath,amssymb,amsfonts}

\def \be {\begin{equation}}
\def \ee {\end{equation}}

\def \bea {\begin{eqnarray}}
\def \eea {\end{eqnarray}}
\def \bal  {\begin{align}}
\def \eal  {\end{align}}
\def \nn {\nonumber}

\def \a {\alpha}

\def \g {\gamma}
\def \G {\Gamma}
\def \d {\delta}

\def \m {\mu}
\def \n {\nu}

\def \vp {\varphi}

\def \o {\omega}
\def \O {\Omega}

\def\bo {\bar{\o}}

\def \p {\partial}

\def\bd{\begin{document}}
\def\ed{\end{document}}
\def\nn{\nonumber}
\def\bea{\begin{eqnarray}}
\def\eea{\end{eqnarray}}
\let\bm=\bibitem
\let\la=\label
\def\hs{\hspace}
\def\ol{\overline}

\def\N{{\cal N}}
\def\sst{\scriptscriptstyle}
\def\thetabar{\bar\theta}
\def\Tr{{\rm Tr}}
\def\one{\mbox{1 \kern-.59em {\rm l}}}

\newcommand{\ttbs}{\char'134}
\newcommand\fverb{\setbox\fverbbox=\hbox\bgroup\verb}
\newcommand\fverbdo{\egroup\medskip\noindent
			\fbox{\unhbox\fverbbox}\ }
\newcommand\fverbit{\egroup\item[\fbox{\unhbox\fverbbox}]}
\newbox\fverbbox
\newcommand{\jhepname}{JHEP}

\title{Hidden conformal symmetry of extremal Kaluza-Klein black hole in four dimensions}

\author{Yong-Chang Huang, Fang-Fang Yuan\\
	Institute of Theoretical Physics, Beijing University of Technology\\
    Beijing 100124, China\\
    E-mail: \email{ychuang@bjut.edu.cn, ffyuan@emails.bjut.edu.cn}
    }	

\received{\today}
\revised{\today}
\accepted{\today}

\abstract{We study the hidden conformal symmetry of four-dimensional extremal Kaluza-Klein black hole. The scalar Laplacian corresponding to the radial equation in the near-region is rewritten in terms of the $SL(2,\mathbb R)$ quadratic Casimir. Using the first law of black hole thermodynamics, this symmetry enables us to obtain the conjugate charges for the CFT side. The real-time correlators are also found to agree with the CFT expectations.}

\begin{document}

\section{Introduction}
The Kerr/CFT correspondence is a powerful duality that relates rotating black holes to two-dimensional conformal field theory (2D CFT) \cite{S01}. Since this gives us a way to get access to more realistic black holes, in particular, the Kerr black hole, rather than supersymmetric ones, it has invigorated very extensive studies since the advent\footnote{For a partial list of recent develepments, see \cite{KC01.1} - \cite{KC01.11}.}. Based on Brown-Henneaux formalism \cite{BH}, i.e., by studying the asymptotic symmetry group, the Kerr/CFT correspondence has been applied to many different cases.

An important issue that perplexed the Kerr/CFT for some time is the microscopic description of non-extremal black holes. %\cite{La} % the non-extremal case of the duality
Remarkably, Castro, Maloney and Strominger uncovered the hidden conformal symmetry of non-extremal Kerr black hole in \cite{CMS}. In a specific limit that is called near-region, it was found that the scalar Laplacian corresponding to the wave equation could be rewritten as the quadratic Casimir of an $SL(2,\mathbb R) \times SL(2,\mathbb R)$ algebra. This was fulfilled by introducing a set of "conformal coordinates" and defining some vector fields as the generators of the algebra. In contrast with the extremal case, the local conformal symmetry in \cite{CMS} was in the solution space of wave equation. The aspects of entropy and absorption cross section have also been discussed there for the non-extremal Kerr black hole. This formalism is quite universal which could be applied to many kinds of black holes as indicated by subsequent works \cite{KC02.1} - \cite{KC03.4}.

In particular, a new set of conformal coordinates was introduced for extremal black holes in \cite{BC07}\footnote{A related work appeared in \cite{Mat}.}. The $SL(2,\mathbb R)$ quadratic Casimir was also obtained. Except the applications in \cite{BC07}, this has been extended to other black holes \cite{KC03.1,KC03.2,KC03.3,KC03.4}. Another new aspect of the hidden conformal symmetry was discovered in the study of Kerr-Newman black hole \cite{CC06}. The usual method based on \cite{CMS} can be used to investigate the wave equation which involves three charges, i.e., mass $M$, angular momentum $J$, and charge $Q$. However, the observation of the authors was that when the quantum number of angular momentum $J$ was taken to zero and a new vector operator was introduced for the charge $Q$, another hidden conformal symmetry emerged, which was called Q-picture description. (In view of this, the former description was named as J-picture.) Certain black holes have been analyzed from this viewpoint \cite{KC04.1,KC04.2}.

In this work, we will explore the hidden conformal symmetry of four-dimensional (4D) extremal Kaluza-Klein black hole using the set of conformal coordinates introduced in \cite{BC07}. Note that the 4D non-extremal Kaluza-Klein black hole has been studied in \cite{Ren} along the line with \cite{CMS}. However, the extremal case is nontrivial as shown in previous works on other black holes.

The organization of the paper is as follows. In the next section, following \cite{BC07}, we reproduce the scalar Laplacian of the radial wave equation by deriving the quadratic Casimir, which is the essential indicator of the hidden conformal symmetry. In section \ref{sec:3}, we relate the first law of black hole thermodynamics to the CFT expression of entropy, and obtain the conjugate charges. In section \ref{sec:4}, the Euclidean correlator is written down and we compare it with the real-time correlator on the gravity side. Other microscopic aspects of the duality are also discussed in these two sections. In particular, besides the extremal case, we determine the form of conjugate charges and real-time correlator for the non-extremal case as complementary to the previous study \cite{Ren}. The conclusions and remarks are included in section \ref{sec:5}.

\section{Hidden conformal symmetry} \label{sec:2}

In this section, we will adapt the conformal coordinates of \cite{BC07} to 4D extremal Kaluza-Klein black hole, and reproduce the scalar wave equation in near-region in the low frequency limit. Some basic formulas of Kaluza-Klein black hole can be found in Appendix \ref{sec:A}.

We start with the Klein-Gordon equation for a massless scalar field in the background of 4D Kaluza-Klein black hole
\begin{eqnarray}
 \frac{1}{\sqrt{-g}}\partial_\mu\left
 (\sqrt{-g}g^{\mu\nu}\partial_\nu\right)\Phi=0\;.
 \end{eqnarray}
By using the ansatz of the wave function as
\begin{eqnarray}
 \Phi=e^{-i\omega t+im\varphi}\Phi(r,\theta)\;
 \end{eqnarray}
 and taking a near-region limit which is dictated by
  \begin{eqnarray}
 \omega\mu\ll 1\;,\;\;\;\omega r\ll 1\; ,
 \end{eqnarray}
we obtain the reduced radial wave equation \cite{Ren}
\begin{align} \label{nerad}
%\begin{split}
&\left[\partial_r\Delta\partial_r
 +\frac{\left(2\mu r_+\omega/\sqrt{1-\nu^2}-am\right)^2}{(r-r_+)(r_+-r_-)}
 -\frac{\left(2\mu r_-\omega/\sqrt{1-\nu^2}-am\right)^2}{(r-r_-)(r_+-r_-)}\right]R(r)\nn\\
&\hspace{6.2cm} =l(l+1)R(r).
%\end{split}
\end{align}
As we will concentrate on the extremal limit $\mu=a$, the equation simplifies further as
\begin{align} \label{radial}
 &\left[\partial_r\Delta\partial_r
+\frac{2(2\mu \omega/\sqrt{1-\nu^2})\left(2\mu r_+\omega/\sqrt{1-\nu^2}-am\right)}{r-r_+}
+\frac{\left(2\mu r_+\omega/\sqrt{1-\nu^2}-am\right)^2}{(r-r_+)^2}
  \right]R(r)\nn\\
&\hspace{8cm}=l(l+1)R(r).
\end{align}
 The conformal coordinates for extremal black holes introduced in \cite{BC07} are
\begin{align} \label{excc 1}
\omega^{+}&=\frac{1}{2}(\alpha_1 t+\beta_1 \varphi-\frac{\gamma_1}{r-r_+}),\\
\omega^{-}&=\frac{1}{2}\left(\exp(2\pi T_L\varphi+2n_Lt)-\frac{2}{\gamma_1}\right),\\
 y&=\sqrt{\frac{\gamma_1}{2(r-r_+)}}\exp(\pi T_L\varphi+n_L t) \label{excc 2}.
\end{align}
Thus the explicit expression for the quadratic Casimir is
\begin{align} \label{casi}
H^2&=\partial_r(\Delta\partial_r)-\left(\frac{\gamma_1(2\pi T_L\partial_t-2n_L\partial_{\varphi})}{A(r-r_+)}\right)^2~~~~~~~~~~\nn\\
&-\frac{2\gamma_1(2\pi T_L\partial_t-2n_L\partial_{\varphi})}{A^2(r-r_+)}(\beta_1\partial_t-\alpha_1\partial_{\varphi})~~~~~~~~~~~~~~
\end{align}
where $A=2\pi T_L\alpha_1-2n_L\beta_1$, and $\Delta=(r-r_+)^2$.
Through the comparison of equations (\ref{radial}) and (\ref{casi}), we have the identifications
\begin{eqnarray}
\alpha_1=0,~~~~~~~~~~\beta_1=-\frac{\gamma_1}{\mu},~~~~~~~~2\pi T_L=1,~~~~~~~~~n_L=\frac{\sqrt{1-\nu^2}}{4\mu},
\end{eqnarray}
and also $A=\frac{\sqrt{1-\nu^2}}{2\mu^2}\g_1$. Note that although the equation about $T_L$ here is obtained by a  detailed comparison, it is the same as that of \cite{Ren} where a derivation was given. The fact that the radial wave equation (\ref{radial}) could be reproduced by the quadratic Casimir in (\ref{casi}) in some sense justifies the universal applicability of the new set of conformal coordinates proposed in \cite{BC07}.

Following the logic of \cite{CMS}, we demonstrate the hidden conformal symmetry of 4D extremal Kaluza-Klein black hole explicitly as below.

Firstly, we introduce the conformal coordinates
\begin{align}
\omega^{+}&=-\frac{\g_1}{2}(\frac{\varphi}{\m}+\frac{1}{r-\m}), \label{CC1}\\
\omega^{-}&=\frac{1}{2}\left(\exp(\varphi + \frac{\sqrt{1-\nu^2}}{2\mu}t)-\frac{2}{\gamma_1}\right),\\
y&=\sqrt{\frac{\gamma_1}{2(r-\m)}}\exp(\frac{\varphi}{2} + \frac{\sqrt{1-\nu^2}}{4\mu}t). \label{CC3}
\end{align}

Then we define locally the vector fields
\begin{align}
H_1&=i\partial_{+}\label{H0}\nn\\
%&=i\frac{4\mu^2}{\g_1 \sqrt{1-\nu^2}}(\partial_t-\frac{\sqrt{1-\nu^2}}{2\mu}\partial_{\varphi})
&=i\frac{2\m}{\g_1}(\frac{2\m}{\sqrt{1-\n^2}}\p t - \p \vp),\\
H_0&=i(\omega^{+}\partial_{+}+\frac{1}{2}y\partial_{y})\nn\\
&=-i\left[(r-\m)\p r + \frac{2\mu}{\sqrt{1-\nu^2}}\p t - \p\vp \right], \\
H_{-1}&=i(\omega^{+2}\partial_{+}+\omega^{+}y\partial_{y}-y^2\partial_{-}) \nn\\
&= i \g_1 \left[\frac{r-\m}{\m} \vp \p r
+\frac{1}{\sqrt{1 - \n^2}} (\vp^2 + \frac{\m (3 \m - 2r)}{(r-\m)^2}) \p t
- \frac{1}{2 \m}(\vp ^2 + \frac{\m ^2}{(r -\mu) ^2}) \p \vp\right]
\end{align}
and
\begin{align}
\overline{H}_1&=i\partial_{-}\label{H10}\nn\\
%&= 2i\exp(-\vp - \frac{\sqrt{1 - \n^2}}{2\m}t)\left[(r-\m) \p r
%+ \frac{\m}{r-\m}(\frac{2(r-2\m)}{\sqrt{1 - \n^2}} \p t + \p \vp)\right] \nn\\
&= i\frac{2}{r-\m}\exp(-\vp - \frac{\sqrt{1 - \n^2}}{2\m}t)\left[(r-\m)^2 \p r
+ \m(\frac{2(r-2\m)}{\sqrt{1 - \n^2}} \p t + \p \vp)\right]\\
\overline{H}_0&=i(\omega^{-}\partial_{-}+\frac{1}{2}y\partial_{y})\nn\\
&= -i\frac{2}{\g_1(r-\m)}\exp(-\vp - \frac{\sqrt{1 - \n^2}}{2\m}t)\left[(r-\m)^2 \p r
+ \m(\frac{2(r-2\m)}{\sqrt{1 - \n^2}} \p t +  \p \vp)\right] + i \frac{2\m}{\sqrt{1 - \n^2}} \p t \\
\overline{H}_{-1}&=i(\omega^{-2}\partial_{-}+\omega^{-}y\partial_{y}-y^2\partial_{+})\label{H1}\nn\\
&= -i \frac{1}{2(r-\m)}\exp(\vp + \frac{\sqrt{1 - \n^2}}{2\m}t)
\left[(r-\m)^2 \p r - \m(\frac{2(r-2\m)}{\sqrt{1 - \n^2}} \p t - \p \vp)\right] \nn\\
&+ i \frac{2}{\g_1^2(r-\m)} \exp(- \vp - \frac{\sqrt{1 - \n^2}}{2\m}t)
\left[(r-\m)^2 \p r + \m(\frac{2(r-2\m)}{\sqrt{1 - \n^2}} \p t + \p \vp)\right]
- i \frac{4\m}{\g_1 \sqrt{1-\n^2}} \p t
\end{align}
each set of which satisfies the $SL(2,\mathbb R)$ algebra
\begin{eqnarray}
 ~~[H_0,H_{\pm1}]=\mp i H_{\pm 1},~~~~~~~~[H_{-1},H_1]=-2iH_0,\\
~[\overline{H}_0,\overline{H}_{\pm1}]=\mp i \overline{H}_{\pm 1},~~~~~~~~[\overline{H}_{-1},\overline{H}_1]=-2i\overline{H}_0 .
\end{eqnarray}
In contrast with the extremal case, we have a free parameter $\g_1$ here.

The quadratic Casimir is
\begin{eqnarray} \label{Cas}
H^2=\bar{H}^2=-H_{0}^2+\frac{1}{2}(H_1H_{-1}+H_{-1}H_{1})=
\frac{1}{4}(y^2\partial_{y}^2-y\partial_{y})+y^2\partial_{+}\partial_{-}.
\end{eqnarray}
By inserting equations (\ref{H0}) - (\ref{H1}) into the formula (\ref{Cas}), we find that the expression of the Casimir is exactly that of the scalar Laplacian in (\ref{radial}).

Some comments about the locality of the vector fields defined in equations (\ref{H10}) - (\ref{H1}) are in order. It can be attributed to the fact that under the periodic identification of the angular coordinate $\vp \sim \vp + 2\pi$, they do not behave periodically\footnote{This argument associated with rotating black holes was originally given in \cite{CMS} in the context of non-extremal Kerr/CFT.}. Since they are not globally defined, we cannot use them to generate new global solutions. The corresponding behaviors of conformal coordinates given in equations (\ref{CC1}) - (\ref{CC3}) can also be shown explicitly as follows
\begin{eqnarray} \label{p.i.}
\omega^{+} \sim \omega^{+} - \pi \frac{\g_1}{\m}, \quad \omega^{-} \sim e^{2\pi} (\omega^{-} + \frac{1}{\g_1}) - \frac{1}{\g_1}, \quad y \sim e^{\pi}y,
\end{eqnarray}
where the tranformation property of $\omega^{+}$ is distinct from that of $\omega^{-}$ and $y$.
This identification is generated by the group element
\begin{eqnarray}
e^{-i 2\pi \overline{H}_0}.
\end{eqnarray}
Accordingly, one factor of the $SL(2,\mathbb R)_L \times SL(2,\mathbb R)_R$ symmetry is spontaneously broken to the $U(1)$ subgroup.

Through the comparison between (\ref{p.i.}) and the transformation behaviors of conformal coordinates in non-extremal \cite{CMS} or extremal Kerr/CFT \cite{BC07}, one may read off the temperature as $T_L=\frac{1}{2\pi}$. It also justifies that only one sector exists for the 4D extremal Kaluza-Klein black hole.
On the other hand, from the relation
\begin{eqnarray}
\omega^{-} \sim e^{-t^-},
\end{eqnarray}
where $t^- = - \varphi - \frac{\sqrt{1-\nu^2}}{2\mu}t$, one finds that an observer at fixed position in the Rindler coordinate $t^-$ will observe a thermal bath of Unruh radiation with temperature $ T_L=\frac{1}{2\pi}$.

The hidden conformal symmetry is peculiar in that it pertains to the solution space of near-region wave equation rather than the geometry. It is exactly this broken $SL(2,\mathbb R)$ symmetry that enables us to investigate other aspects of the correspondence in the next two sections\footnote{Note that the essence of the original (extremal) Kerr/CFT \cite{S01} is the realization of Virasoro algebra from an enhancement of the rotational $U(1)$ isometry. It also relies on Brown-Henneaux formalism \cite{BH}.}.

\section{Microscopic description} \label{sec:3}

To be complete, we start with the non-extremal story. The microsopic entropy of 4D Kaluza-Klein black hole is \cite{Ren}
\be
S = \frac{\pi^2}{3}(c_LT_L+c_RT_R) =2\pi\frac{\mu}{\sqrt{1-\nu^2}}(\mu+\sqrt{\mu^2-a^2})
\ee
where
\be \label{tem}
c_L = c_R = \frac{12a\m}{\sqrt{1-\nu^2}},\quad T_L=\frac{r_++r_-}{4\pi a},\quad T_R=\frac{r_+-r_-}{4\pi a}.
\ee
This agrees with the macrocopic Bekenstein-Hawking area law.

In what follows, we determine the conjugate charges from the first law of thermodynamics.
On the CFT side, the entropy is
\be \label{encft}
\delta S=\frac{\d E_L}{T_L} + \frac{\d E_R}{T_R},
\ee
while on the black hole side, we have
\be
\delta S=\frac{1}{T_H} \d M - \frac{\Phi_H}{T_H} \d Q - \frac{\O_H}{T_H} \d J
\ee
which is the first law.
By comparing these two equations, we get the conjugate charges \cite{BC04}
\begin{align}
\delta E_L &= \frac{(2M^2 - Q^2) M}{J} \delta M - \left( \frac{M^2 Q}{J} - \frac{Q^3}{2 J} \right) \delta Q,
\nonumber\\
\delta E_R &= \frac{(2M^2 - Q^2) M}{J} \delta M - \frac{M^2 Q}{J} \delta Q - \delta J.
\end{align}
However, as for our specific case, it is convenient and necessary to express them via the parameters $\m, \n$, and $a$, instead of the physical parameters $M, Q$, and $J$. Note that the discussions concerning the hidden conformal symmetry of
4D Kaluza-Klein black hole as in the above section and in \cite{Ren} all use these parameters.

Based on the formulas in Appendix \ref{sec:A}, the variations of $M, Q$, and $J$ lead to
\begin{align}
\d M = \frac{1}{1-\n^2}(\frac{2-\n^2}{2}\d \m + \frac{\m \n}{1-\n^2}\d \n), \\
\d Q = \frac{1}{1-\n^2}\left(\n\d \m + \m \frac{1+\n^2}{1-\n^2}\d \n\right), \\
\d J = \frac{1}{\sqrt{1-\n^2}}(a\d \m + \frac{a\m\n}{1-\n^2}\d \n + \m \d a).
\end{align}
Using the expressions of angular velocity and electric potential in \ref{potential}, we obtain
\be
T_H \d S = \frac{r_+^2}{r_+^2 + a^2} \d \m + \frac{\n}{2(1-\n^2)} \sqrt{\m^2 - a^2} \d \n - \frac{a\m}{r_+^2+a^2} \d a.
\ee
Recall that $T_H=\frac{\sqrt{1-\nu^2}}{2\pi}\frac{\sqrt{\mu^2-a^2}}{r_+^2+a^2}$, so the variation of entropy can be written as
\be
\d S = \frac{\pi}{\sqrt{1-\n^2}}\left(\frac{2r_+^2}{\sqrt{\m^2 - a^2}}\d \m + \frac{\n}{1-\n^2}(r_+^2 + a^2) \d \n
- \frac{2a\m}{\sqrt{\m^2 - a^2}} \d a \right).
\ee
Then via the comparison of this equation and the CFT fromula (\ref{encft}), and with the help of (\ref{tem}), finally the conjugate charges read as
\begin{align}
\d E_L &= \frac{2\m}{\sqrt{1-\nu^2}} \d \m + \frac{\m^2 \n}{(1-\n^2)^{3/2}} \d \n \\
\d E_R &= \frac{r_+^2 - 2a\sqrt{\mu^2-a^2}}{a\sqrt{1-\nu^2}} \d \m
+ \frac{\m\n\sqrt{\mu^2-a^2}} {a(1-\n^2)^{3/2}}(r_+ - a) \d \n - \frac{\m}{\sqrt{1-\nu^2}} \d a.
\end{align}
With the identifications
\begin{align} \label{iden 1}
\d \m &=  \o, \hspace{3ex}\d \n = e,\hspace{3ex}\d a = m, \\
\o_L &= \frac{2\m}{\sqrt{1-\nu^2}} \o, \hs{3ex}
\o_R = \frac{r_+^2 - 2a\sqrt{\mu^2-a^2}}{a\sqrt{1-\nu^2}} \o - \frac{\m}{\sqrt{1-\nu^2}} m,  \\
q_L&=q_R=\d \n=e,\\
\m_L &= - \frac{\m^2 \n}{(1-\n^2)^{3/2}},
\hs{3ex}\m_R= - \frac{\m\n\sqrt{\mu^2-a^2}} {a(1-\n^2)^{3/2}}(r_+ - a)\label{iden 2}
\end{align}
we have
\be
\d E_L=\o_L-q_L\m_L, \hs{3ex}\d E_R=\o_R-q_R\m_R.
\ee

Then we turn to the extremal case, and the quantities corresponding to the extremal case will be denoted with a prime. %to distinguish.
Now the microscopic entropy comes entirely from the left sector and takes the form
\be
S' = \frac{\pi^2}{3}c_LT'_L =2\pi\frac{\mu^2}{\sqrt{1-\nu^2}}
\ee
where $T'_L = \frac{1}{2\pi}$.

Note that the CFT formula (\ref{encft}) is now modified to
\be
\delta S'=\frac{\d E'_L}{T_L'}.
\ee

Proceeding as the above discussion, the variation of the extremal entropy is
\be
\d S' = \frac{4\pi \m}{\sqrt{1-\n^2}} \d \m + \frac{2\pi \m^2 \n}{(1-\n^2)^{3/2}} \d \n.
\ee
Thus the conjugate charge is immediately found as
\be
\d E_L' = \frac{2 \m}{\sqrt{1-\n^2}} \d \m + \frac{\m^2 \n}{(1-\n^2)^{3/2}} \d \n.
\ee
If we identify
\begin{align} \label {exiden 1}
\d \m &=  \o, \hspace{3ex}\d \n = e,\hspace{3ex}\d a = m, \\
\o'_L &= \frac{2\m}{\sqrt{1-\nu^2}} \o, \hs{3ex}
\\
q'_L&=\d \n=e, \\
\m'_L &= - \frac{\m^2 \n}{(1-\n^2)^{3/2}}\label{exiden 2},
\end{align}
it follows that
\be
\d E'_L=\o'_L-q'_L\m'_L.
\ee
Note that the identifications (\ref{exiden 1}) - (\ref{exiden 2}) are the same as the left sector part of those for the non-extremal case (\ref{iden 1}) - (\ref{iden 2}).

To sum up, we have derived the conjugate charges for both the 4D non-extremal and extremal Kaluza-Klein black holes.

\section{Real-time correlator}  \label{sec:4}

In this section, we will compare the Euclidean correlator on the CFT side to the real-time correlator on the black hole side, and find the agreement. A general study of the real-time correlators in the Kerr/CFT correspondence can be found in \cite{BC01}. For the discussion about the extremal case, see \cite{BC07}.

In a 2D CFT, the two-point functions of the primary operators are determined by the
conformal invariance. The Euclidean correlator is related to the retarded correlator as
\be \label{GER} G_E(\o_{L,E}, \o_{R,E}) = G_R(i\o_{L,E},
i\o_{R,E}), \quad \o_{L,E} , \o_{R,E} >0. \ee
At finite temperature, $\o_{L,E}$ and $\o_{R,E}$ take discrete
values of the Matsubara frequencies \be \o_{L,E} =  2 \pi m_L T_L,
\quad \o_{R,E} =  2 \pi m_R T_R. \ee

For an operator of dimensions $(h_L,h_R)$,
%charges $(q_L,q_R)$ at temperatures $(T_L,T_R)$ and chemical potentials $(\m_L, \m_R)$,
the momentum space Euclidean correlator is given by \cite{MS97}
 \bea \label{GE}
G_E(\o_{L,E}, \o_{R,E}) &\sim& T_L^{2 h_L-1}  T_R^{2 h_R-1} e^{i
\frac{\bo_{L,E}}{2T_L}} e^{i \frac{\bo_{R,E}}{2T_R}}\nn\\
&&\cdot\G(h_L + \frac{\bo_{L,E}}{2 \pi T_L})\G(h_L -
\frac{\bo_{L,E}}{2 \pi T_L}) \G(h_R + \frac{\bo_{R,E}}{2 \pi
T_R})\G(h_R - \frac{\bo_{R,E}}{2 \pi T_R}), \eea where \be
\bo_{L,E}= \o_{L,E} - i q_L \m_L, \quad \bo_{R,E}= \o_{R,E} - i
q_R \m_R. \ee

On the gravity side, for a scalar field in a black hole background, the prescription for
two-point real-time correlators was originally proposed in \cite{Son05}. We will follow \cite{BC01} where
this technique has been adapted to the Kerr black hole.
%For the scalar wave function satisfying the ingoing boundary condition at the black hole horizon,
%its asymptotic behavior is \be \phi \sim A r^{h-1}+B r^{-h}. \ee
%Then the two-point retarded correlator is just \bea G_R &\sim & \frac{B}{A}
%\eea

%For the charged scalar scattering off the Kerr-Newman black hole,
%its asymptotic behavior looks like (\ref{chargedscalar}. Therefore
%its retarded Green's function is just
% \bea
%G_R&\sim&\frac{\G(1-2h)}{\G(2h-1)}\frac{\G\left(h+i\frac{\o_L-q_L\m_L}{2\pi
%T_L}\right)\G\left(h+i\frac{\o_R-q_R\m_R}{2\pi T_R}\right)}
% {\G\left(1-h+i\frac{\o_L-q_L\m_L}{2\pi T_L}\right)\G\left(1-h+i\frac{\o_R-q_R\m_R}{2\pi
% T_R}\right)}
 % \eea
 %with the identification (\ref{identification1}-\ref{identification3}). This
%is in good match with the CFT prediction (\ref{GER}).

For the 4D Kaluza-Klein black hole, by introducing
\be
z=\frac{r-r_+}{r-r_-},
\ee
the radial equation (\ref{nerad}) can be written as
\be
(1-z)z\p^2_z R(z)+(1-z)\p_z R(z)+\left(\frac{A_1}{z}+A_2+\frac{A_3}{1-z}\right) R(z)=0,
\ee
where
\bea
A_1&=& \frac{\left(2\mu r_+\omega/\sqrt{1-\nu^2}-am\right)^2}{(r_+-r_-)^2}, \nn\\
A_2&=&- \frac{\left(2\mu r_-\omega/\sqrt{1-\nu^2}-am\right)^2}{(r_+-r_-)^2},\nn\\
A_3&=&-l(l+1).
\eea
The central information comes from the asymptotic behavior
\be \label{exas}
R(r) \sim A r^{h-1}+B r^{-h} \ee
where
$h$ is the conformal weight \be
h=l+1. \ee
The retarded Green's function is then
\bea
G_R&\sim&\frac{\G(1-2h)}{\G(2h-1)}\frac{\G\left(h+i\frac{\bo_L}{2\pi
T_L}\right)\G\left(h+i\frac{\bo_R}{2\pi T_R}\right)}
 {\G\left(1-h+i\frac{\bo_L}{2\pi T_L}\right)\G\left(1-h+i\frac{\bo_R}{2\pi
 T_R}\right)}.
  \eea
With the identifications (\ref{iden 1}) - (\ref{iden 2}), the Green's function is obviously in agreement with the CFT expression (\ref{GE}).

For the extremal case, only the left sector contributes. However, as noted in \cite{BC07}, the asymptotic behavior
is nontrivially different from the above discussion as indicated below. As in the above section, the extremal
quantities will be denoted with a prime to distinguish from the non-extremal ones.

On the CFT side, the momentum space Euclidean correlator is now given by
 \bea \label{exGE}
G_E(\o'_{L,E}, \o'_{R,E}) &\sim& {T'_L}^{2 h'_L-1}
e^{i\frac{\bo'_{L,E}}{2T'_L}}\G(h'_L + \frac{\bo'_{L,E}}{2 \pi T'_L})\G(h'_L -
\frac{\bo'_{L,E}}{2 \pi T'_L}), \eea
where \be
\bo'_{L,E}= \o'_{L,E} - i q'_L \m'_L. \ee

The extremal radial equation (\ref{radial}) can be written as
\begin{eqnarray}
[\partial_r \Delta\partial_r+\frac{C}{r-r_+}
+\frac{D^2}{(r-r_+)^2}]R(r)=l(l+1)R(r)~~~~~~~~~~~~~~
\end{eqnarray}
where
\begin{eqnarray}
C&=& 2(2\mu \omega/\sqrt{1-\nu^2})\left(2\mu r_+\omega/\sqrt{1-\nu^2}-am\right)  \nn \\
D&=& 2\mu r_+\omega/\sqrt{1-\nu^2}-am.
\end{eqnarray}
Introduce $x=\frac{-2iD}{r-r_+}$, then the equation leads to
\begin{eqnarray}
\frac{d^2R}{dx^2}+(\frac{\frac{1}{4}-m_s^2}{x^2}+\frac{k}{2}-\frac{1}{4})R(x)=0
\end{eqnarray}
where
\begin{eqnarray}
k=i2\mu \omega/\sqrt{1-\nu^2},~~~~~~~~~~~~~m_s^2=\frac{1}{4}+l(l+1).
\end{eqnarray}
Different from the non-extremal case (\ref{exas}), the asymptotic behavior is now
\begin{eqnarray}
R\sim C_1r^{-h'}+C_{2}r^{1-h'}
\end{eqnarray}
where $h$ is the conformal weight
\begin{eqnarray}
h'=\frac{1}{2}+m_{s}=\frac{1}{2}+\sqrt{\frac{1}{4}+l(l+1)}.
\end{eqnarray}
%\begin{eqnarray}
%C_1=-\frac{\Gamma(1-2m_s)}{\Gamma(\frac{1}{2}-m_s-k)}D,~~~~~~~~~~C_2=\frac{\Gamma(1+2m_s)}{\Gamma(\frac{1}{2}+m_s-k)}D
%\end{eqnarray}
%where $D$ is a constant.
Thus the retarded Green function is
\begin{eqnarray}
G_R\sim\frac{C_1}{C_2}\propto\frac{\Gamma(1-2h')\Gamma(h'-k)}{\Gamma(2h'-1)\Gamma{(1-h'-k)}}.
\end{eqnarray}
With the identifications (\ref{exiden 1}) - (\ref{exiden 2}) for the 4D extremal Kaluza-Klein black hole, we find the agreement with the Euclidean correlator on the CFT side (\ref{exGE}).

\newcommand\bit{\noindent$\bullet$ \verb}
\section{Conclusions}  \label{sec:5}

It has been shown in \cite{BC07} that in contrast with the non-extremal case, to study the hidden conformal symmetry of extremal black holes, a new set of conformal coordinates should be introduced. In this paper, we extended this to study the 4D extremal Kaluza-Klein black hole. The scalar Laplacian corresponding to the radial equation in the near-region can be rewritten in terms of the $SL(2,\mathbb R)$ quadratic Casimir. Based on the relation between the
first law of black hole thermodynamics and the CFT expression of entropy, we have obtained the conjugate charges for both the 4D non-extremal and extremal Kaluza-Klein black holes. The real-time correlators also agreed with the CFT expressions. Obviously, the fermionic and vector cases of the scattering and real-time correlators can be investigated analogously which we omitted here.

Some possible extentions of this work are as follows. Firstly, it may be interesting to find a relation with the holographic Q-picture description proposed in \cite{CC06}, whose main point has been sketched in the introduction. Just as the Kerr-Newman black hole, the 4D Kaluza-Klein black hole has three kinds of charges. However, all the discussions here used the new parameters ($\m, \n$, and $a$) instead of the physical parameters ($M, Q$, and $J$). The procedure introduced in \cite{CC06}, especially taking the limit of the wave equation, may need to be slightly modified for our specific case. Secondly, the hidden conformal symmetry of 5D rotating Kaluza-Klein black hole \cite{5DKK.1,5DKK.2} may deserve to study. In particular, the results of the extremal case should be consistent with those in \cite{5DKC}, where the asymptotic symmetry group of this black hole was explored. Finally, a physical explanation may be found for the free parameter in conformal coordinates which exists for the extremal black holes\footnote{See equations (\ref{excc 1}) - (\ref{excc 2}). Note that $\g_1$ is a free parameter which is absent in the non-extremal case.}. One may anticipate its relation
%It is reasonable to relate it
with the special properties of near horizon geometry.

Although the discovery of hidden conformal symmetry is a breakthrough for the non-extremal Kerr/CFT correspondence,
a more systematic approach is still expected. Other than the continuous appearance of new ideas, many corners and aspects of the whole picture remain to be explored.

\acknowledgments

This work was supported by National Natural Science Foundation of China (No. 10875009) and by Beijing
Natural Science Foundation (No. 1072005).

\appendix

\section{Four-dimensional Kaluza-Klein black hole}\label{sec:A}

We review the basics of four-dimensional Kaluza-Klein black hole in this appendix. The Kaluza-Klein black hole is an exact solution of the following Einstein-Maxwell-dilaton action with $\a = \sqrt{3}$
\be
S = \frac{1}{16\pi}\int d^4x\sqrt{-g}\Big[R -2(\nabla\phi)^2 -e^{-2\alpha\phi}F^2\Big].
\ee
It can be obtained from dimensional reduction of the boosted five-dimensional Kerr solution to four dimensions.
The metric is \cite{KK1} - \cite{KM}
\bea
ds^2 &=& -\frac{\Delta -a^2\sin^2\theta}{B\Sigma}dt^2 -2a\sin^2\theta\frac{Z}{B\sqrt{1 -\nu^2}}dtd\varphi \nn \\
&& +\Big[B(r^2+a^2) +a^2\sin^2\theta \frac{Z}{B}\Big]\sin^2\theta d\varphi^2 \\
&& +\frac{B\Sigma}{\Delta}dr^2 +B\Sigma d\theta^2 \, , \nn
\eea
where
\be
\Delta = r^2 -2\mu r +a^2 \, , \qquad \Sigma = r^2 +a^2\cos^2\theta \, , \qquad
 Z = \frac{2\mu r}{\Sigma} \, , \qquad B = \sqrt{1 +\frac{\nu^2Z}{1 -\nu^2}} \, .
\ee \\
The dilaton field is $\phi = -\big(\sqrt{3}/2\big)\ln B$, and the components of gauge potential are
\be
A_t = \frac{\nu Z}{2(1 -\nu^2)B^2} \, , \qquad
A_{\varphi} = -\frac{a\nu Z\sin^2\theta}{2\sqrt{1 -\nu^2}B^2} \, .
\ee
The physical mass $M$, the charge $Q$, and the angular momentum $J$ are expressed through the mass parameter $\mu$, boost parameter $\nu$, and specific angular momentum $a$ as follows
\bea
M = \mu \Big[1 +\frac{\nu^2}{2(1 -\nu^2)}\Big] \, , \qquad\qquad \\
Q = \frac{\mu\nu}{1 -\nu^2} \, , \qquad J = \frac{\mu a}{\sqrt{1 -\nu^2}} \, ,
\eea
and the locations of the horizons are at
\be
 r_\pm=\mu\pm\sqrt{\mu^2-a^2}\;.
\ee
The angular velocity and electric potential of the event horizon are, respectively, given by
\bea \label{potential}
\Omega_H = \frac{a\sqrt{1 -\nu^2}}{r^2_+ +a^2}, \,\qquad
\Phi_H =  \frac{Qr_+ (1 -\nu^2)}{r^2_+ +a^2} =
\frac{\nu}{2} \, .
\eea
Hawking temperature and entropy of the black hole are
\bea
 T_H&=&\frac{\sqrt{1-\nu^2}}{2\pi}\frac{\sqrt{\mu^2-a^2}}{r_+^2+a^2}\;,\nonumber\\
 S&=&2\pi\frac{\mu}{\sqrt{1-\nu^2}}(\mu+\sqrt{\mu^2-a^2})\;.
 \eea
 In this work, we will focus on the extremal limit $\mu=a$ when the entropy is
 \bea
 S(T_H=0)=\frac{2\pi\mu^2}{\sqrt{1-\nu^2}}\;.
 \eea

\end{document}